# An Oscillating - Magnet Watt Balance

## H. Ahmedov

TÜBİTAK, UME, National Metrology Institute of Turkey

E-mail:haji.ahmadov@tubitak.gov.tr

**Abstract**

We establish the principles for a new generation of simplified and accurate watt balances in which an oscillating magnet generates Faraday's voltage in a stationary coil. A force measuring system and a mechanism providing vertical movements of the magnet are completely independent in an oscillating magnet watt balance. This remarkable feature allows to establish the link between the Planck constant and a macroscopic mass by a one single experiment. Weak dependence on variations of environmental and experimental conditions, weak sensitivity to ground vibrations and temperature changes, simple force measuring procedure, small sizes and other useful features offered by the novel approach considerable reduce the complexity of the experimental setup. We formulate the oscillating-magnet watt balance principle and establish the measurement procedure for the Planck constant. We discuss the nature of oscillating-magnet watt balance uncertainties and give a brief description of the Ulusal Metroloji Enstitüsü ( UME ) watt balance apparatus.

1. Introduction

In 1975 Kibble described the principles of the first moving-coil watt balance [1,2], a two-part experiment which links the electrical and mechanical Sİ units by indirect conversion of electrical power into mechanical power. The watt balance principle and macroscopic electrical quantum effects: the Josephson effect and the quantum Hall effect [3, 4] established the link between a macroscopic mass and the Planck constant, the fundamental constant of the microworld [ 5 ]. This link provides a potential route to the redefinition of the kilogram [6, 7], the last base unit which is still defined by a manmade object, the international prototype of the kilogram.

In most moving-coil watt balances the Ampere's force law and the Faraday's law of induction are tested separately. In a weighing phase the gravitational force on a mass is balanced by the Lorentz force on a current-carryng coil in a magnetic field while in a moving phase, the magnetic field induces Faraday's voltage in the moving coil. When combined, these phenomena state the equality of electrical and mechanical powers. Measurement accuracy of the two–part watt balance experiment is limited by variations in the environmental and experimental conditions between moving and weighing phases. The need to quantify these variations at the level of several parts in $10^9$ considerably complexifies the experiment. High sensitivity to ground vibrations and temperature changes, strong dependence on variations of environmental and experimental conditions, non-linear magnetic effects and alignment issues are some examples of such complications.

In addition to the conventional two-experiment mode, the Bureau International des Poids et Mesures ( BIPM ) watt balance can operate in a simultaneous scheme, where weighing and moving phases are carried out in a one single experiment [8]. In the experiment the force comparator is kept fixed, but an extension system, driven by an electrostatic motor in the suspension, allows the coil to move while it is hanging from the force comparator. The electrostatic motor may cause the emergence of forces not related to the watt balance principle. It is possible to reduce these undesirable forces if to move the coil with a constant velocity. However the use of uniform movement leads to a problem of another kind. In the coil in addition to an induced voltage an resistive voltage appears as a consequence of nonzero resistance of the wire. Attempts to separate the constant induced voltage and the resistive voltage with an accuracy sufficient for the realization of the kilogram have not yet led to success [9,10].

Existing experiments differ in the way the coil is moved and guided during the dynamic phase. In the National Physical laboratory ( NPL ) a large balance beam resting on a knife-edge was used in the weighing phase and the movement of the coil in the moving phase was generated by tilting the balance beam [11-13]. The National Research Council ( NRC ) continues the NPL watt balance experiment. Although they made some important modifications to the apparatus, many of the parts remains as originally developed by NPL [14,15]. In the National İnstitute of Standartds and Technology ( NİST ) watt balance the rotation of the large balance wheel leads to a purely vertical movement of the coil [16-18]. In the first Federal Institute of Metrology ( METAS )  experiment the coil was disconnected from the balance in the moving phase and moved by a separate mechanical system [19]. In the new METAS watt balance experiment a highly constrained machanical guiding system ( 13 hinge translation stage ) is used to ensure pure vertical movements of the coil [20]. A distinctive feature of the Laboratoire National de Metrologie et d'Essais (LNE) experiment is that the force comparator is moved together with the coil by a precision guiding stage [21,22]. The Measurement Standards Laboratory (MSL) is pursuing the idea of using a twin pressure balance as the force transducer for the weighing phase and an oscillatory coil movement for the moving phase [23,24]. In the Korea Research İnstitute of Standards and Science ( KRİSS )  experiment a piston and cylinder assembly is used as the motion guiding stage to minimize parasitic motion of the coil  [25]. Detailed description of existing watt balance experiments can be found in review articles [5,7, 26 - 28 ].

Once the kilogram has been redefined, the watt balances will become the principal methods by which an individual National Measurement Institute (NMI) can make an independent determination of the SI unit of mass and thereby contribute to the maintenance of national and international mass scales [29]. Excessive complexity of existing watt balances not only restrict the accuracy of the Planck constant measurements but represents a serious drawback for primary realization of the kilogram in NMI also. NPL have published the theory and some designs for a new generation of simplified moving-coil watt balances. The technique allows the construction of balances which are insensitive to the alignment requirements previously thought necessary for succesful operation [30].

The complexity of moving coil watt balances is largely due to the fact that the coil accepts active participation both at measurement of the Lorentz force and at generation of the Faraday's voltage. In this paper we propose the theory and the basic design for a new generation of watt balances with a stationary coil and a moving magnet. In contrary to the traditional moving coil watt balances a force measuring system and a mechanism which displaces the magnet vertically are completly independent in an oscillating magnet watt balance. The oscillating magnet generates an alternating Faraday voltage in the coil. Since the oscillating Faraday's voltage and the constant resistive voltage are of different nature simultaneous testing of the Faraday's law of induction and the Ampere's force law becomes possible.

In moving-coil watt balances ground vibrations is a limiting factor in the measurement of the Planck constant [31-33]. In the oscillating magnet approach ground vibrations are eliminated by a continuous averaging procedure over the magnet oscillation cycles. The duration of an oscillating magnet watt balance experiment depends on the oscillation frequency and amplitude and on the structure of ground vibrations. The results of vibration measurements conducted in the UME watt balance laboratory suggest that at the amplitude in one millimeter and at frequences more than one Hertz the duration of the experiment should be within 100 seconds to suppress the influence of apparatus vibrations on the measurement accuracy of the Planck constant. At such small durations variations of the magnetic field caused by temperature changes becomes inessential for the Planck constant measurement accuracy. Weak dependence of the Planck constant on ground vibrations and temperature changes not only reduces the size of the apparatus but also allows the experimental setup to operate at normal laboratory conditions without the use of complex vibration isolation and temperature control systems.

In moving-coil watt balances the impact of a weighing current on the field of a magnet is eliminated by using of the forward and reverse currents in the coil [13]. This procedure requires Lorentz force measurements in two opposite directions. In oscillating magnet watt balances due to the unified approach to the watt balance principle Lorentz force measurements in the downward direction is sufficient to establish the link between a macroscopic mass and the Planck constant. This feature simplifies the force measuring procedure and the coil suspension assembly.

The UME oscillating-magnet watt balance project was initiated at the second half of 2014 and presently the manufacturing of the first trial version is in progress.

Measurements on the trial version are expected to be performed during the first half of 2016. At this stage we expect to measure the Planck constant with accuracy of several parts in $10^7$. An improved version would be designed and constructed based on the information obtained from the trial one. Final measurements are planned to be performed during 2017 and expected result is to reach uncertainty level less than $5 \cdot 10^{-8}$.

The paper is organized as follows. In Section 2 we describe the structure of unmeasured mechanical and electrical powers and establish an oscillating - magnet watt balance principle. In Section 3 we establish a measuring procedure for the Planck constant and propose two methods for simultaneous measurements of the Faraday's voltage and the Ampere's current. In Section 4 we analyse vibrational, thermal and alignment uncertainties and give a brief description of the UME oscillating-magnet watt balance apparatus.

## 2. The oscillating- magnet watt balance principle

In principle every process which transforms electrical energy into mechanical one can be used to establish a link between a macroscopic mass and the Planck constant. However, every direct energy conversion suffers from energy losses, which would need to be quantified at the level of several parts in $10^9$, which is very demanding. The peculiarity of the watt balance principle is that it provides indirect conversion of electrical energy into mechanical energy without energy losses. This principle is based on two basic laws of the quasistationary electrodynamics. Consider a physical system, consisting of a coil and a magnet that are in motion, and assume that an electric current $J$ flows through the coil. According to the Ampere's force law the magnet generates in the coil a Lorentz force

$$\vec{F}(t) = \vec{G}(t)J(t) \quad (2.1)$$

and a Lorentz torque

$$\vec{K}(t) = \vec{H}(t)J(t) \quad (2.2)$$

Here $\vec{G}$ and $\vec{H}$ are factors which clarify the interaction. They depend on the structure of the magnetic field and the geometry of the coil. According to the Faraday's law of induction the coil acquires an induced voltage

$$\frac{h}{h_{90}}\mathcal{E}(t) = \vec{G}(t)\vec{V}(t) + \vec{H}(t)\vec{\Omega}(t) \quad (2.3)$$

Here $\vec{V}$ and $\vec{\Omega}$ are linear and angular velocities of the coil relative to the magnet. The presence of the Planck constant $h$ and the conventional value $h_{90}$ of the Planck constant in the Faraday's law is due to the fact that since 1990 the electrical units of voltage and resistance are based on the conventional values of the Josephson and the Klitzing constants [34].

By combining the Ampere's force law and the Faraday's law of induction we establish the oscillating-magnet watt balance equation

$$\frac{h}{h_{90}}\langle j \rangle \delta \mathcal{E} = \langle F_z \rangle (1 + \frac{\delta G_z}{\langle G_z \rangle})\delta V_z + \langle \vec{F}_\perp \rangle \delta \vec{V}_\perp + \langle \vec{K} \rangle \delta \vec{\Omega} \quad (2.4)$$

Here $F_z$ is the component of the Lorentz force along the direction of gravitational acceleration and $\vec{F}_\perp$ is the projection of the Lorentz force into a horizontal plane. The link between a macroscopic mass and the Planck constant is achieved by comparing a gravitational force and the mean value $\langle F_z \rangle$ which is the average of $F_z$ over the duration of the experiment. The original purpose of watt balances is to determine the Planck constant by means of the existing SI kilogram standard. For the redefinition of the kilogram unit the Planck constant should be measured with accuracy better that 2 part in $10^8$. After achievement of such an accuracy a value of the Planck constant will be fixed and watt balances will become the principal method for the realization the kilogram.

Since in an oscillating magnet watt balance experiment physical quantities have dynamic character it is useful to work with the mean deviation

$$\delta f = f - \langle f \rangle \quad (2.5)$$

We first rewrite the Faraday's law of induction in terms of mean deviations by omitting all terms which do not affect the accuracy of the experiment. Then using the Ampere's force law we find the mean value of the G-factor. In this way we arrive at the oscillating magnet watt balance equation (2.4).

Figure 1 illustrates an oscillating-magnet watt balance experiment. A coil is immersed in the air gap of an magnetic circuit which oscillates in a vertical direction. A closed type magnetic circuit with two opposing magnetic rings is the most practical solution for a watt balance realising the kilogram [35-37]. Such a magnetic circuit produces rotational and up-down symmetric field in the air gap. These symmetries allows us to introduce a symmetric axis and a symmetric plane for the magnetic circuit. Their intersection define the center of the magnetic circuit. This geometric center coincides with the mass center. Similar considerations are valid also for the coil which have the form of a cylindrical ring. We assume that symmetric axes of the magnet and the coil are paralel to the direction of gravitational acceleration which is depicted by a unit vector $\vec{n}$. Then the force and the torque factors are determined by a displacement vector $\vec{X}$ which represents the deviation of the coil center from the magnet center

$$\vec{G} = G_z \vec{n}, \quad \vec{H} = [\vec{X}, \vec{G}] \quad (2.6)$$

where

$$\delta G_z = \langle G_z \rangle \left( \delta \zeta + \delta f(X_z) \right) \quad (2.7)$$

The magnetic field and the geometry of the coil changes with temperature. These changes are described by a function $\zeta(t)$. A functions $f$ describes vertical inhomogeneity of the magnetic field in the air gap. Due to the up-down symmetry the function $f(X_z)$ contains only even degrees of the vertical displacement

$$\delta f(X_z) = \Lambda_2(a) \frac{\delta X_z^2(t)}{a^2} + \Lambda_4(a) \frac{\delta X_z^4(t)}{a^4} + \cdots \quad (2.8)$$

Here $a = \sqrt{2\langle (\delta X_z)^2 \rangle}$ is the oscillation amplitude. Note that the G-factor may also have a horizontal inhomogeneity which is defined by the horizontal displacement $\vec{X}_\perp$. However, the impact of the horizontal inhomogeneity on the Planck constant is negligible.

So far we have assumed that the magnetic circuit produces ideal radial magnetic field and the coil is of ideal cylindrical form. Small deviations from ideal conditions may violate the up-down symmetry and as a result odd powers of $\delta X_z$ may appear in the function $\delta f$. Small deviations from ideal conditions may affect the H-factor also. In general case it is determined by so called electrical centers of the coil and the magnet [16, 20]. This effect can be taken into account by means of the replacement $\vec{X} \to \vec{X} + \vec{b}$ in ( 2.6 ) where $\vec{b}$ is a vector which magnitude depends on the degree of imperfection of the coil and the magnetic circuit. When the symmetric axes of the coil and the magnet do not concide with the direction of gravitational acceleration there appear horizontal components for the G factor ( see appendix A ).

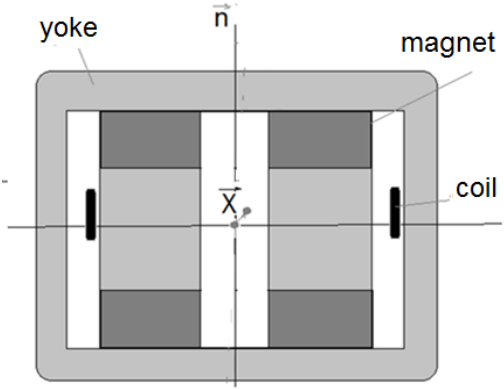

Figure 1. The cross-section of the magnetic circuit and the coil

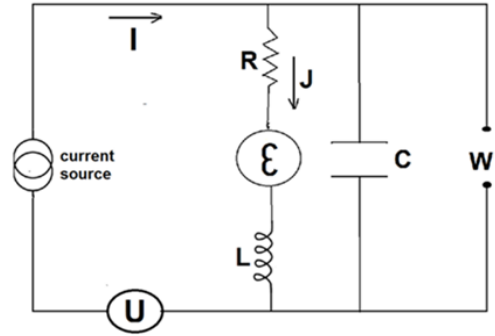

Figure 2. The oscillating-magnet watt balance electrical circuit.

The unified approach to the watt balance principle causes the voltage $W$ at the output of the coil does not coincide with the Faraday's voltage $\mathcal{E}$. Figure 2 illustrates the electrical circuit which describes electrical processes in the experiment. We have added an additional voltage source U into the electrical circuit. The reason for doing this will be clear in the next Section. Neglecting time variations of the capacitance and the inductance we find the following relation

$$\delta \mathcal{E} = (1 - \omega^2/\omega_0^2)\delta W + \langle R \rangle \langle C \rangle \delta \dot{W} - \langle I \rangle \delta R - \langle R \rangle \delta I - \langle L \rangle \delta \dot{I} \quad (2.9)$$

Here $\omega_0 = 1/\sqrt{\langle L \rangle \langle C \rangle}$ is the resonance frequency of the electrical circuit. Because of the smallness of the reactance impedance we have replaced the second derivative $\delta \ddot{W}$ in the above relation with $-\omega^2 \delta W$. Here $\omega$ is the fundamental frequency of the oscillating magnet. It is assumed that all other harmonics are small enough to be taken into account.

Simultaneous realization of the watt balance principle gives rise to a difference between the measured current $\langle I \rangle$ and the Ampere's current $\langle J \rangle$ also. However, this effect is too small to affect the accuracy of the experiment.

Taking into account all the above we may rewrite the oscillating magnet watt balance equation (2.4) in the following form

$$\langle I \rangle \delta W \frac{h}{h_{90}} = \langle F_z \rangle (1 + \omega^2/\omega_0^2 + \delta f(X_z))\delta V_z + \delta P \quad (2.10)$$

In the oscillating watt balance experiment measured quantities are the following: mean values of the input current $\langle I \rangle$ and the Lorentz force $\langle F_z \rangle$, mean deviations of the output voltage $\delta W$ and the coil relative velocity $\delta V_z$, the resonance frequency $\omega_0$ of the electrical circuit, the fundamental frequency $\omega$ of the magnet oscillation and the $\Lambda_2$ parameter which describes the vertical quadratic inhomegeneity. All other quantities are included in a function $\delta P$. This function is a total unmeasured power and is equal to the sum of the electrical power (A.1) and the mechanical power (A.2).

3. **Measurement procedure**

A magnetic circuit commits a vertical oscillation with a fundamental frequency $\omega$ and a fundamental period $T$. We divide the duaration $\tau = NT$ of the experiment on $2N$ intervals

$$t_k = \frac{T}{2}k, \quad k = 0,1,\ldots,2N-1 \quad (3.1)$$

and assume that the magnetic circuit takes minimum and maximum positions at even k and odd k respectively. We average the output voltage and the relative coil velocity over the oscillation half period

$$\langle W \rangle_k = \frac{2}{T}\int_{t_k}^{t_{k+1}} W(t)\, dt. \quad (3.2)$$

Applying this integration to the watt balance equation (2.10) we obtain the Planck constant $h = \bar{h} + \delta\bar{h}$. The measured part of $h$ is

$$\bar{h} = q\, h_{90} \frac{\langle F_z \rangle}{\langle I \rangle} \frac{1}{2N} \sum_{k=0}^{2N-1} \frac{\langle \delta V_z \rangle_k}{\langle \delta W \rangle_k} \quad (3.3)$$

A factor $q$ describes the vertical inhomogeneity of the G-factor and the coil reactance empedance

$$q = 1 + \omega^2/\omega_0^2 + \frac{\Lambda_2(a)}{6} + \frac{\Lambda_4(a)}{20} \quad (3.4)$$

Inhomogenities associated with odd powers of $\delta X_z$ vanishes due to the averaging procedure over $2N$ oscillation cycles. The magnetic field of the coil induces eddy currents in the moving magnetic circuit. The oscillating magnetic field of the eddy currents in turn generates an additional dynamic force in the coil which is proportional to the relative velocity of the coil. This nonlinear effect is described in the appendix B. Due to the smallness of the eddy current coeffient (B.4) the averaging procedure over oscillating cycles makes the influence of eddy currents on the Planck constant measurement accuracy negligible small.

Detailed description of the UME magnetic circuit and the coil is given in Section 4. Using their metrological characteristics we find

$$\Lambda_2(a) = \Lambda_2(a_0) \frac{a^2}{a_0^2}, \quad \Lambda_2(a_0) \cong 1 \cdot 10^{-6}, \quad a_0 = 1mm \quad (3.5)$$

At the amplitude $a_0$ the quartic inhomogeneity is negligible. By decreasing the oscillation amplitude it is possible to make the quadratic inhomogeneity negligible also. However, the use of such small amplitudes may lead to complications of another kind. First of all interferometers with higher resolutions will be required for displacement measurements. Another difficulty may arise from ground vibrations. Preliminary analysis shows that amplitudes near to $a_0$ are most suitable for watt balance experiments. At such amplitudes the measurement of the $\Lambda_2$ coefficient with accuracy of several parts in $10^2$ will be sufficient. The magnetic circuits commits an almost oscillatory motion $\delta X_z$ with a fundamental frequency $\omega$. The $\Lambda_2$ parameter is linked with the cubic function $\delta X_z^3$ which have the resonance at $3\omega$ frequency. We apply the Fourier transformation to the Faraday's law of induction (2.3) and obtain

$$\Lambda_2(a_0) \frac{a^2}{a_0^2} = \left( F_\omega[\delta V_z] \frac{F_{3\omega}[\delta \mathcal{E}]}{F_\omega[\delta \mathcal{E}]} - F_{3\omega}[\delta V_z] \right) / F_{3\omega}\left[ \frac{\delta X_z^2 \delta V_z}{a^2} \right] \quad (3.6)$$

Here $F_\omega[f]$ shows the $\omega$ - Fourier component of a function $f(t)$. The third harmonic $F_{3\omega}[\delta V_z]$ of the vertical velocity may influence the measurement accuracy. This harmonic can be reduced by a suitable choice of a moving mechanism. Since the right hand side of (3.6) does not depend on the amplitude $a$ we may reach the desired accuracy for the $\Lambda_2(a_0)$ paramer by increasing of the ratio $a/a_0$. Note that it is possible to determine the quadratic inhomogeneity by using of the Ampere's force law too. This can be achieved by comparing Lorentz forces measured at different vertical displacements.

The unmeasured part of the Planck constant

$$\delta\bar{h} = h_{90}\frac{1}{\langle I \rangle}\frac{1}{2N}\sum_{k=0}^{2N-1}\frac{\langle \delta P \rangle_k}{\langle \delta W \rangle_k} \quad (3.7)$$

may be represented in the form

$$\delta\bar{h} = \delta\bar{h}_e + \delta\bar{h}_m \quad (3.8)$$

Here an electrical uncertainty $\delta\bar{h}_e$ and a mechanical uncertainty $\delta\bar{h}_m$ are determined by the unmeasured electrical power and the unmeasured mechanical power respectively.

The use of an oscillatory movement instead of a uniform movement provides the possibility to separate the Faraday's voltage from the resistive voltage at room temperatures. Indeed from (2.9) we observe that mean deviation of the Faraday's voltage does not contain the coil resistive voltage $\langle I \rangle\langle R \rangle$. The unmeasured electrical power is given by (A.1) where current fluctuations have the following form

$$\delta I = \delta I_0 + \frac{\langle R \rangle}{\langle R_0 \rangle}(\delta W + \delta U) \quad (3.9)$$

Here $I_0$ is the current produced by a current source. The impact of the Faraday voltage on the electrical circuit is determined by the ratio of the coil resistance and the current source internal resistance $\langle R_0 \rangle$. We have added an additional voltage $\delta U$ to the electrical circuit to weaken the effect of the Faraday's voltage on the current source. The averaging procedure over oscillation cycles eliminates the reactance empedance related to the first derivative of the voltage. In the appendix A we show that thermal fluctuations of the coil and the current source is inessential in oscillating magnet watt balance experiments. As a result the electrical uncertainty of the apparatus is of the form

$$\delta\bar{h}_e = \frac{\langle R \rangle}{\langle R_0 \rangle}(1 + \frac{1}{2N}\sum_{k=0}^{2N-1}\frac{\langle \delta U \rangle_k}{\langle \delta W \rangle_k}) \quad (3.10)$$

By suitable choice of the voltage $\delta U$ one can make this uncertainty sufficiently small.

It is possible to take into account the electrical uncertainty without the use of the compensating voltage source. This can be achieved by estimating the resistances of the current source and the coil. In this case we add $\delta\bar{h}_e$ to the measured part $\delta\bar{h}$ of the Planck constant. Changes of the resistances with temperature may bring complications to electrical measurements. However by reducing the coil resistance and using current sources with high internal resistance temperature effects may be done insignificant.

## 4. Description of the apparatus

The total unmeasured mechanical energy has three sources: the vibrational energy, the energy of thermal deformations of the magnet and the coil, and the energy that develops from rotations and linear motions in a horizontal plane. These energies give rise to vibrational, thermal and alignment uncertainties. Using (A.7) we get the following expression for the vibrational uncertainty

$$\delta h_m^{(v)} = h_{90}\, \vec{\Gamma}\, \langle\vec{\Phi}\rangle, \quad \vec{\Gamma} = \frac{1}{2N}\sum_{k=0}^{2N-1}\frac{\langle \delta \vec{V}^{(v)}\rangle_k}{\langle \delta V_z\rangle_k} \quad (4.1)$$

Here $\langle\vec{\Phi}\rangle$ is the deflection of the Lorentz force from the vertical axis and $\delta\vec{V}^{(v)}$ is the velocity of horizontal vibrations. Analysis of the $\vec{\Gamma}$ factor shows that for a given measurement duration the vibrational uncertainty is inversely proportional to the oscillation amplitude and the oscillation frequency. Ground vibration measurements carried out in the UME watt balance laboratory give the following estimation

$$|\vec{\Gamma}| \approx 5 \cdot 10^{-6}.\ N = 200,\ a = 1mm \quad (4.2)$$

To make the alignment uncertainty ( 4.6 ) reasonable small the deflection angle should be less than $10^{-3}$ radian. Taking into account the smallness of the deflection angle we conclude that vibrations of apparatus will no affect the Planck constant accuracy.

The number of cycles $N$ and the oscillation frequency determines the duaration $\tau$ of the watt balance experiment. Using (4.1) and ( 4.2 ) we conclude that at the oscillation frequencies in the range $1\ Hz - 2\ Hz$ the duration of the experiment is about $100\ s$. At such small durations temperature changes of the G -factor has linear time dependence

$$\delta\zeta = \sigma\left(t - \frac{\tau}{2}\right) \quad (4.3)$$

Here $\sigma$ is a parameter depending on physical properties ( such as density, heat capacitance, thermal conductivity ) of a magnetic circuit and a thermodynamic state of an environment. In the appendix A we show that variations of the G-factor caused by such temperature changes will no influence the accuracy of the Planck constant. This circumstances leads to a significant size reduction of the apparatus, in particular of the magnet. In the UME watt balance the closed type radial symmetric magnetic circuit is used. Figure 3 shows a cross-sectional view of the magnetic circuit. The magnet shown here is a variation of the NIST4 magnet system design [35] with one notable difference: The sizes in all spatial directions are reduced three times. The yoke is made from armco pure iron and the permanent magnets from SmCo. From the general theory of magnetic circuits we know that the distribution of the magnetic field remains invariant under the rescaling of the size. This property enables to describe properties of the UME magnetic circuit based on results obtained for the NIST4 magnet. The UME magnetic circuit produces a radial field of about $B = 0,55\ T$ in the centre of the air gap. The width of the air gap is 10 mm and the relative variation of the field in the vertical direction is smaller than $10^{-4}$ over 25 mm in the centre of the air gap. These characteristics of the magnetic circuit determine cross sectional dimensions of the coil which has the mean diameter of 144 mm.

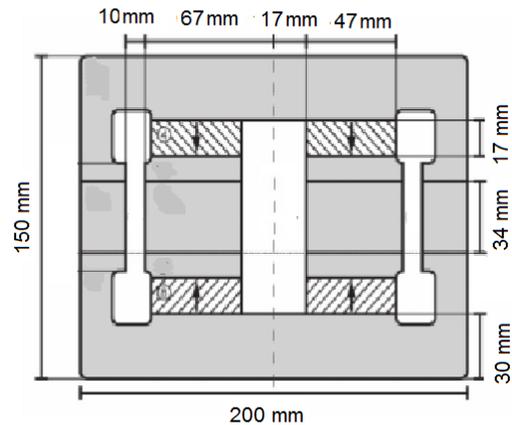

Figure 3. Cross-sectional view of the UME magnetic system

To simplify electrical measurements it is useful to use coils with low resistance. Coil resistance is directly proportional to the total length of the wire and is inversely proportional to the cross-sectional area of the wire. By means of the Biot-Savart law we find the total lenght of the coil wire $L = \langle G_z \rangle / B$. The Ampere's force law links the force factor with a macroscopic mass

$$\langle G_z \rangle = Mg / \langle I \rangle \quad (4.4)$$

As a result we obtain the following expression for the coil resistance

$$\langle R \rangle = \frac{2\rho}{\pi dS} \left( \frac{Mg}{B \langle I \rangle} \right)^2 \quad (4.5)$$

Here $\rho$ is the electrical resistivity of the copper, $S$ is the cross-sectional area of the coil and $d$ is the mean diameter of the coil. The mean diameter and the cross-sectional area of the coil are restricted by the geometry of the air gap. The maximum cross-sectional area that the coil may have is $S = 200\ mm^2$. For a given mass the coil resistance is inversely proportional to the square of the Ampere's current. Electrical currents are measured as voltaghe drop across a calibrated resistor. By means of two 100 Ω Wilkins-type standards placed in parallel high accuracy measurements of the current $\langle I \rangle = 16\ mA$ is possible [12]. From the equation ( 4.5 ) we obtain $\langle R \rangle = 400\ \Omega$ and $\langle R \rangle = 100\ \Omega$ for masses $M = 1\ kg$ and $M = 0.5\ kg$ respectively. By using several 100 Ω Wilkins-type standards placed in parallel it is possible to reduce the resistance of the coil even more.

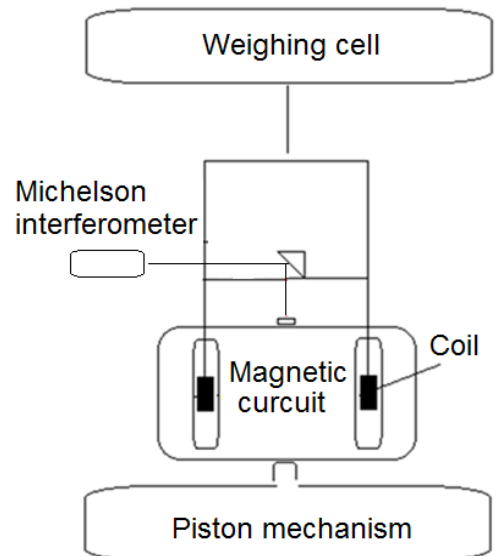

Figure 4. Configuration of the UME watt balance apparatus

The general view of the UME oscillating magnet watt balance is shown in Figure 4. The apparatus has radial symmetry. The outer diameter and the height of the experimental setup is 500 mm and 700 mm respectively. The force measuring system consists of a weighing cell and a coil suspended from it which is immersed in the magnetic circuit. The coil is connected to its support frame using three identical rods, which are equally spaced around the circumference of the coil former. İnhomogeneity of the magnetic field in the gap and eddy currents give rise to a dynamic Lorentz force in the coil. In the previous section and in the appendix B we have shown that these effects are of the order of several parts in million. In force measurements we are planning to use the Mettler Toledo AX5006 mass comparator which electrical weighing range is between + 11 g and - 1 g. Since the electrical weighing range is much greater than the Lorentz force oscillation amplitude the appearance of dynamic force will not lead to difficulties at force measurements.

The magnetic field of the coil magnetises the yoke. As a result the magnetic field of the yoke will depend on the current flowing through the coil. In the oscillating magnet watt balance the Faraday's voltage and the Lorentz force are measured in the same magnetic field. Therefore the dependence of the force factor on the current will not affect the Planck constant measurement accuracy. The oscillating-magnet approach allows to establish the direct link between a kilogram standard and the Planck constant. The gravitational force of the mass is compared with the vertical component of the Lorentz force directed downward. This feature of the oscillating-magnet watt balance leads to a significant simplification of the force measuring process and the coil suspension system.

A very important topic is alignment because although an imperfectly aligned apparatus can provide very reproducible results, they are systematically wrong. The alignment uncertainty of the apparatus is given by

$$\delta h_m^{(a)} = h_{90} \left( \langle \vec{\Phi} \rangle \langle \vec{\Psi} \rangle + \frac{\vec{a} \langle \vec{K} \rangle}{a \langle F_z \rangle} \right) \quad (4.6)$$

Here $\langle \vec{\Psi} \rangle$ is the deflection of the relative coil velocity from the vertical direction and $\langle \vec{K} \rangle$ is the torque in the coil. It is natural to suggest that the relative angular velocity oscillates with the same frequency as the linear velocity. In the above expression $\vec{a}$ represents the amplitude of the angular oscillation. From equations of motions one can show that the amplitude of angular fluctuations is proportional to the amplitude of linear fluctuations. Independence of the alignment uncertainty from the oscillation amplitude is important since otherwise the use of small amplitudes may increase the contribution of the Lorentz torque to the apparatus uncertainty. In the UME watt balance apparatus a piston mechanism is used to provide vertical movements of the magnet. Piston motion is controlled by a servo-motor. In this way almost oscillatory motion can be achieved. Mechanical constraints are used to decrease the angular amplitude $\vec{a}$ and the deflection angle $\langle \vec{\Psi} \rangle$.

The relative velocity of the coil is determined by a Michelson interferometer. A corner cube is installed in the middle of the magnet upper surface and the beam splitter is installed at the center of the coil support frame.

## 5. Conclusion

We have proposed the theory and the basic design for the oscillating magnet watt balance which provides the link between a macroscopic mass and the Planck constant in a most natural way without any complications. High accuracy, small sizes, an ability to operate at normal laboratory conditions without use of complex vibration isolation system and temperature control system, simplicity of the force measuring system, short duration of the experiment are major advantages of the oscillating magnet approach. The only source of uncertainty of the apparatus is related to alignments. Alignment issues can be solved by imposing mechanical constraints on the mechanism providing vertical motions of the magnet.

As the main objective of the article is an establishment of the general principles for the new system we tried to avoid technical detail. This applies primarily to the theory of electrical measurements and to the analysis of apparatus vibrations. We have described in general terms two approaches that can simultaneously measure the Faraday's voltage and the Ampere's current at room temperatures. A more detailed presentation of these techniques will be given in a forthcoming paper. Using general considerations we have established the link between vibration of the apparatus and the duration of the experiment. More detailed analysis of the vibrational uncertainty by taking into account both stationary and nonstationary ground vibrations will be given later.

**Acknowledgment**

I would like to thank my collegues from the Physical and the Mechanical departments who have provided assistance for this project. I would like to express special thanks to T.C. Öztürk from the voltage laboratory for her valuable assistance in the development of the theory of simultaneous electrical measurements and to C. Birlikseven from the time and the frequency laboratory for his effective assistance in the preparation of various experiments and data processing. Finally the author gratefully thank S. Schlamminger from NIST for useful discussions.

### Appendix A. Mechanical and electrical powers

The total unmeasured power $\delta P$ is the sum of the electrical power

$$\delta P_e = \langle I \rangle (\langle I \rangle \delta R + \langle R \rangle \delta I + \langle L \rangle \delta \dot{I} - \langle R \rangle \langle C \rangle \delta \dot{W}) \quad (A.1)$$

and the machanical power

$$\delta P_m = \langle F_z\rangle\delta\zeta\delta V_z + \langle\vec{F}_\perp\rangle\delta\vec{V}_\perp + \langle\vec{K}\rangle\delta\vec{\Omega} \quad (A.2)$$

The projection of the Lorentz force and the coil relative velocity into a horizontal plane is given by

$$\langle\vec{F}_\perp\rangle = \langle F_z\rangle\left[\langle\vec{\Phi}\rangle,\vec{n}\,\right] \quad (A.3)$$

and

$$\delta\vec{V}_\perp = \delta V_z[\langle\vec{\Psi}\rangle,\vec{n}] + [\delta\vec{V}^{(v)},\vec{n}\,] \quad (A.4)$$

Here $\langle\vec{\Phi}\rangle$ and $\langle\vec{\Psi}\rangle$ are deflections of the Lorentz force and the relative coil velocity from the vertical axis. $\langle\vec{\Phi}\rangle$ is equal to the sum of deflection angles of the magnet and the coil symmetric axes. Square brackets represent the vector product operation and $\delta\vec{V}^{(v)}$ characterizes horizontal vibrations of the apparatus. Inserting (A.3) and (A.4) into (A.2) we obtain

$$\delta P_m = \delta P_m^{(t)} + \delta P_m^{(v)} + \delta P_m^{(a)} \quad (A.5)$$

where

$$\delta P_m^{(t)} = \langle F_z\rangle\delta\zeta\delta V_z \quad (A.6)$$

is the thermal power

$$\delta P_m^{(v)} = \langle F_z\rangle\langle\vec{\Phi}\rangle\delta\vec{V}^{(v)} \quad (A.7)$$

is the vibrational power and

$$\delta P_m^{(a)} = \langle F_z\rangle\langle\vec{\Phi}\rangle\langle\vec{\Psi}\rangle\delta V_z + \langle\vec{K}\rangle\delta\vec{\Omega} \quad (A.8)$$

is the power which includes both rotations and linear motions in the horizontal plane. These Powers determines the mechanical uncertainty of the apparatus

$$\delta\bar{h}_m = \delta h_m^{(v)} + \delta h_m^{(t)} + \delta h_m^{(a)} \quad (A.9)$$

The thermal uncertainty of the experimental setup is

$$\delta h_m^{(t)} = h_{90}\frac{1}{2N}\sum_{k=0}^{2N-1}\frac{\langle\delta\zeta\,\delta V_z\rangle_k}{\langle\delta V_z\rangle_k} \quad (A.10)$$

where the function $\delta\zeta$ is given by ( 4.3 ). Tests conducted for the UME magnetic circuit show that the parameter σ is of the order of $10^{-9}$ Hertz. İt is not difficult to verify that (B.1) is much smaller than $10^{-9}h_{90}$. Thermal fluctuations of the coil and the current source are treated in a similar fashion.

## Appendix B. Eddy current effect

In the presence of eddy currents the force factor mean deviation (2.7) looks as follows

$$\delta G_z = \langle G_z \rangle \left( \delta \zeta + \delta f(X_z) + \Lambda_{ed} \frac{\delta V_z}{a\omega} \right) \quad (B.1)$$

The eddy current effect is proportional to the relative coil velocity where the eddy current coefficient is of the form

$$\Lambda_{ed} = \sigma \left(\frac{\mu_0}{4\pi}\right)^2 \frac{a\omega}{Mg} \int_D d^3\vec{x}\ \vec{Q}(\vec{x}) \cdot \vec{Q}(\vec{x}) \quad (B.2)$$

Here $D$ is the region of the yoke, $\sigma$ is the conductivity of the yoke ( the iron ) and

$$\vec{Q}(\vec{x}) = \int_{D_c} d^3\vec{y}\ \frac{\vec{n}\cdot(\vec{x}-\vec{y})}{|\vec{x}-\vec{y}|^3} \vec{j}(y) \quad (B.3)$$

where $D_c$ is the region of the coil and $\vec{j}$ is the current dencity flowing through the coil. Rough numerical estimate results in

$$\Lambda_{ed} < 2 \cdot 10^{-5} \quad (B.4)$$

We have used the following data: the mass $M = 1\ kg$, the frequency $f = 1 Hz$, the thikness of the coil is 8 mm and the heght of the coil is 25 mm.